\def\spose#1{\hbox to 0pt{#1\hss}}

\def\multleft#1{\hbox to size{\vbox {\halign {\lft{##}\cr #1}}\hfill}\par}
\def\multright#1{\hbox to size{\vbox {\halign {\rt{##}\cr #1}}\hfill}\par}

\def\today{\ifcase\month\or January\or February\or March\or April\or May\or
      June\or July\or August\or September\or October\or November\or December\fi
      \space\number\day, \number\year}
\def\asec{$^{\prime\prime}$}

\newcommand{\Msolar}{\mbox{\,$\rm M_{\odot}$}}

\def\H2{\hbox{H$_{2}$}}

\newcommand{\ltsim}{\mbox{{\raisebox{-0.4ex}{$\stackrel{<}{{\scriptstyle\sim}}
$}}}}

\documentclass{mn2e}

\usepackage{times}
\usepackage{amssymb}
\usepackage{epsfig}

\begin{document}
\hsize=6truein
          
\title[Lyman break galaxies]
{The luminosity function, halo masses and stellar masses of 
luminous Lyman-break galaxies at redshifts $\bf 5<z<6$}

\author[R.J.~McLure et al.]
{R. J. McLure$^{1}$\thanks{Email: rjm@roe.ac.uk}, M. Cirasuolo$^{1}$, 
J. S. Dunlop$^{1,2}$, S. Foucaud$^{3}$, O. Almaini$^{3}$\\
\footnotesize\\
$^{1}$SUPA\thanks{Scottish Universities Physics Alliance} Institute
for Astronomy, University of Edinburgh, Royal Observatory, Edinburgh EH9 3HJ\\
$^{2}$Department of Physics \& Astronomy, University of British Columbia, 6224
Agricultural Road, Vancouver, BC, V6T 1Z1, Canada\\
$^{3}$School of Physics and Astronomy, University of Nottingham,
University Park, Nottingham NG7 2RD} 

\maketitle

\begin{abstract}
We present the results of a study of a large sample of luminous ($z_{AB}^{\prime}<26$) Lyman
break galaxies (LBGs) in the redshift interval $4.7<z<6.3$, selected
from a contiguous 0.63 square degree area covered by the
UKIDSS Ultra Deep Survey (UDS) and the Subaru XMM-Newton Survey (SXDS). 
Utilising the large area coverage and the excellent available
optical+nearIR data, we use a photometric redshift analysis to derive
a new, robust, measurement of the bright end ($L\geq L^{\star}$) of
the UV-selected luminosity function at high redshift. When combined with literature studies of the
fainter LBG population, our new sample provides improved
constraints on the luminosity function of redshift $5<z<6$ LBGs over
the luminosity range 0.1L$^{\star}$$\ltsim$ L $\ltsim10$L$^{\star}$. A maximum
likelihood analysis returns best-fitting Schechter function parameters
of $M_{1500}^{\star}=-20.73\pm0.11, \phi^{\star}=0.0009\pm0.0002\, {\rm
Mpc}^{-3}$ and $\alpha=-1.66\pm0.06$ for the luminosity function at $z=5$, and
$M_{1500}^{\star}=-20.04\pm0.12, \phi^{\star}=0.0018\pm0.0005\,{\rm Mpc}^{-3}$
and $\alpha=-1.71\pm0.11$ at $z=6$. In addition, an analysis of the
angular clustering properties of our LBG sample demonstrates that 
luminous $5<z<6$ LBGs are strongly clustered
($r_{0}=8.1^{+2.1}_{-1.5}h_{70}^{-1}$Mpc), and are consistent with the
occupation of dark matter halos with masses of
$\simeq10^{11.5-12.0}\Msolar$. Moreover, by stacking the available multi-wavelength
imaging data for the high-redshift LBGs it is possible to place useful
constraints on their typical stellar mass. The results of this
analysis suggest that luminous LBGs at $5<z<6$ have an
average stellar mass of $\log_{10}({\rm{M}/\Msolar})=10.0^{+0.2}_{-0.4}$, consistent
with the results of the clustering analysis assuming plausible values
for the ratio of stellar to dark matter. Finally, by combining our
luminosity function results with those of the stacking analysis we
derive estimates of $\simeq1\times10^7 \Msolar$ Mpc$^{-3}$ and
$\simeq4\times10^6 \Msolar$ Mpc$^{-3}$ for the stellar mass density at
$z\simeq5$ and $z\simeq6$ respectively.
\end{abstract}


\section{INTRODUCTION}
Accurately determining the global properties of the high-redshift 
galaxy population is a powerful method of constraining
current models of galaxy evolution, and identifying the sources responsible for reionisation. 
Large, statistical samples of high-redshift galaxies can be identified
efficiently using two complementary photometric techniques. Firstly,
Lyman-alpha emitters (LAEs) can be selected via deep imaging
with narrow-band filters centred on the redshifted Lyman-alpha
emission line (Hu, McMahon \& Cowie 1999). Alternatively, so-called
Lyman-break galaxies (LBGs) can be selected from deep broad-band
photometry using the Lyman-break, or ``dropout'', technique pioneered by
Guhathakurta, Tyson \& Majewski (1990).

When combined with deep multi-wavelength photometry and spectroscopic
follow-up, narrow-band selection of LAEs is an efficient technique 
for producing large samples of high-redshift galaxies within narrow
redshift intervals, free from significant levels of contamination. By
exploiting the wide-field imaging capabilities of Suprime-Cam on
Subaru, several recent studies have investigated the number densities,
luminosity functions and clustering properties of LAEs at $z=5.7$ and
$z=6.6$ (e.g. Ouchi et al. 2008; Shimasaku et al. 2006; Taniguchi et
al. 2005). Moreover, at present, the highest-redshift galaxy with
spectroscopic confirmation ($z=6.96$; Iye et al. 2006) was identified using the
narrow-band technique (see Stark et al. 2007a for candidate LAEs at $z\geq8$).
However, although LAE studies have many advantages, they are 
constrained by the fact that any individual study can only survey 
a comparatively small cosmological volume and that only a minority of
LBGs at high redshift appear to be strong LAEs (e.g. Shapley et al. 2003).
Therefore, although the selection function of the Lyman-break
technique is more difficult to quantify, because it includes both LAEs
and non-LAEs, it should provide a more complete census of the high-redshift galaxy population.

Consequently, many studies in the recent literature have exploited
both ground-based and Hubble Space Telescope (HST) data to study the
properties of high-redshift galaxies selected via the Lyman-break technique.
In particular, due to its unparallelled point-source sensitivity, deep HST imaging has
allowed studies of the high-redshift luminosity function to reach
$\ltsim 0.1L^{\star}$, and has therefore been crucial to
constraining both the normalization and faint-end slope of the
luminosity function at $3<z<6$ (e.g. Dickinson et al. 2004; Malhotra
et al. 2005; Oesch et al. 2007; Bouwens et al. 2006, 2007, 2008; Reddy et al. 2007). 
In terms of best-fitting Schechter function parameters the results of
these studies have described a fairly consistent picture, with the majority finding a very
steep faint-end slope, typically in the range
$-1.5<\alpha<-1.8$ (although see Sawicki \&  
Thompson 2006 for a different result at $z\simeq4$), and little
evidence for significant evolution in either the faint-end slope or
normalization at $z\geq3$.

Although studies based on deep HST imaging have been successful at
exploring the faint end of the high-redshift luminosity function,
their small areal coverage means that they are not ideally suited
to accurately constraining the bright end of the luminosity function. 
To address this issue studies based on shallower, but wider area, ground-based imaging
have been important (e.g. Ouchi et al. 2004a; Shimasaku et al. 2005;
Yoshida et al. 2006; Iwata et al. 2007). To date, although the
existing wide-area, ground-based optical imaging surveys
have been very useful for constraining the bright end of the LBG
luminosity function in the redshift range $3<z<5$, it has been
difficult to successfully extend ground-based studies to higher redshift.
The reason for this is that without the benefits of HST spatial
resolution to exclude low-redshift contaminants (i.e red galaxies at
$z\simeq1$ and ultra-cool galactic stars), selecting clean samples of
$z>5$ LBG candidates requires near-infrared
photometry\footnote {The use of two medium-band $z-$filters to quantify the UV slope of a
small sample of 12 redshift $5.6<z<6.2$ candidates in the Subaru Deep
Field by Shimasaku et al. (2005) is one notable exception.} to confirm a
relatively blue spectral index long-ward of Lyman alpha.
Unfortunately, due to the lack of wide area near-infrared detectors,
until very recently it was not possible to obtain suitably deep
near-infrared imaging over areas commensurate with the existing wide-area optical surveys.

However, with the advent of deep near-infrared imaging over a 0.8 square~degree
field provided by the UKIDSS Ultra Deep Survey (UDS) this fundamental
issue is now being addressed. In a previous paper (McLure et al. 2006) we
combined the early data release of the UDS with optical imaging from
the Subaru/XMM-Newton survey (SXDS) to explore the properties of the
most luminous ($z^{\prime}\leq25$) galaxies at $z\geq5$.
In this paper we extend this previous study by using new, deeper, $JK$
imaging of the UDS field to study the properties of a much larger
sample of $z^{\prime}\leq26$ LBGs in the redshift range $4.7<z<6.3$.
The specific aim of this study was to use the large, contiguous area of the UDS
to provide an improved measurement of the bright end of
the UV-selected luminosity function at $z=5$ and $z=6$, and to measure the
clustering properties, and thereby halo masses, of the luminous LBG population at $z\geq5$.

The structure of the paper is as follows. In
Section 2 we briefly describe the properties of the near-infrared and
optical data-sets used in this study. In Section 3 we describe
our initial selection criteria and our photometric redshift
analysis. In Section 4 we describe our adopted technique for
estimating the LBG luminosity function. In Section 5 we present our
luminosity function results, and the results of combining our
ground-based data with existing constraints derived from deep HST imaging data.
In Section 6 we report the results of our study of the clustering
properties of the luminous LBG population in the redshift range
$5<z<6$ and link the LBG population to that of the underlying dark
matter halos. In Section 7 we provide an estimate of the typical
stellar mass for luminous $5<z<6$ LBGs based on a stacking analysis of
the available multi-wavelength imaging data. In Section 8 we use
this information to derive an estimate of the stellar mass function at
$z=5$ and $z=6$ and compare our results with the predictions of recent 
semi-analytic models of galaxy formation. In Section 9 we summarise our main conclusions. Throughout
the paper we adopt the following cosmology: $H_{0}=70$ km\,\,s$^{-1}$Mpc$^{-1}$, $\Omega_{m}=0.3$, 
$\Omega_{\Lambda}=0.7$, $n_s=1.0$, $\sigma_8=0.9$. All magnitudes are
quoted in the AB system (Oke \& Gunn 1983).

\section{The data}
The Ultra Deep Survey (UDS) is the deepest of five
near-infrared surveys currently underway at the UK InfraRed Telescope
(UKIRT) with the new WFCAM imager (Casali et al. 2007) which together 
comprise the UKIRT Infrared Deep Sky Survey (UKIDSS; Lawrence et
al. 2007). The UDS covers an area of 0.8 square degrees centred on
RA=02:17:48, Dec=$-$05:05:57 (J2000) and is already the deepest, large area,
near-infrared survey ever undertaken. The data utilized in this paper
were taken from the first UKIDSS data release (DR1;  Warren et al. 2007), 
which included $JK$
imaging of the entire UDS field to $5\sigma$ depths of $J=23.9,
K=23.8$ (1.6\asec diameter apertures). The UKIDSS DR1 became publicly 
available to the world-wide astronomical community in January 2008 and can be downloaded from 
the WFCAM Science Archive. \footnote{http://surveys.roe.ac.uk/wsa/}

The UDS field is covered by a wide variety of deep, multi-wavelength
observations ranging from the X-ray through to the radio (see
Cirasuolo et al. 2008 for a recent summary). However, for the
present study the most important multi-wavelength observations are the deep optical
imaging of the field taken with Suprime-Cam (Miyazaki et
al. 2002) on Subaru as part of the Subaru/XMM-Newton Deep 
Survey (Sekiguchi et al. 2005). The optical imaging consists of 5 
over-lapping Suprime-Cam pointings, and covers an area of $\simeq
1.3$~square degrees. The whole field has been imaged in the
$BVRi^{\prime}z^{\prime}$ filters, to typical $5\sigma$ depths of 
$B=27.9$, $V=27.4$, $R=27.2$, $i^{\prime}=27.2$ and $z^{\prime}=26.2$
(1.6\asec diameter apertures). The reduced optical imaging of the
SXDS is now publicly available
\footnote{http://www.naoj.org/Science/SubaruProject/SDS/} and full
details of the observations, data reduction and calibration procedures
are provided in Furusawa et al. (2008). The high-redshift galaxies
investigated in this study were selected from a contiguous area of
0.63 square degrees (excluding areas contaminated by bright stars and
CCD blooming) covered by both the UDS near-infrared and SXDS optical
imaging.

\begin{figure}
\centerline{\psfig{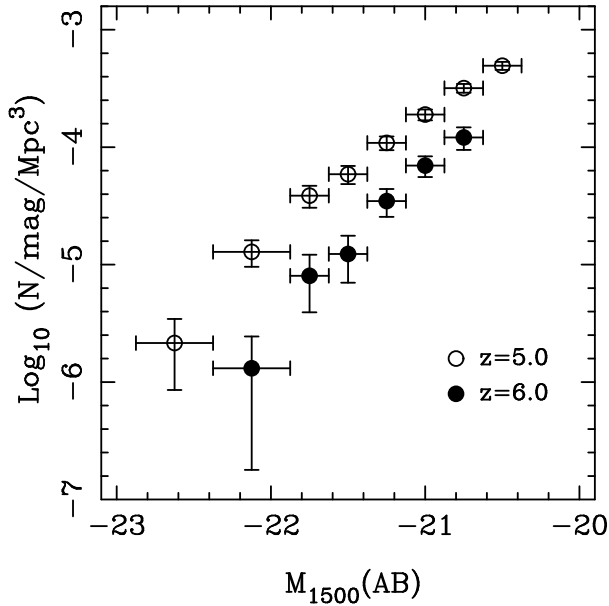}}
\caption{The $V/V_{max}$ estimates of the bright end of the UV-selected luminosity
function in two redshift intervals centred on $z=5$ \& $z=6$. The
absolute UV magnitudes have been calculated at a rest-frame wavelength
of 1500\AA. The horizontal error bars indicate the width of the magnitude bins adopted
(either $\Delta m=0.25$ or $\Delta m=0.5$ depending on
signal-to-noise). The vertical error bars indicate the uncertainty due
to simple poisson statistics. The faintest bin in the luminosity
function estimate is set by our adopted magnitude limit of
$z^{\prime}=26$. The bright limit is set by the requirement that 
the brightest bin should contain the equivalent of more than one object.}
\end{figure}

\section{High redshift candidate selection}

One of the principle motivations for this study was to exploit the
unique combination of areal coverage and optical+nearIR data available
in the UDS/SXDS to study the high-redshift galaxy population without
recourse to strict dropout/LBG selection criteria. The reasoning
behind this is to allow the luminosity function to be studied 
over a relatively wide range of redshifts from a single sample, and 
to reduce as much as possible the strong bias towards young, blue star-forming galaxies
inherent to traditional Lyman-break selection. Although single-colour
selection techniques (e.g. $V$-drop, $i-$drop) have proven to be highly
successful at isolating high-redshift galaxies, because of the need to
adopt fairly strict colour-cut criteria (e.g. $i-z>1.3$ for
$z\geq5.5$), it is at least possible that a significant population of 
old/redder, perhaps more massive, galaxies which
marginally fail to satisfy these strict criteria could be excluded
(e.g. Dunlop et al. 2007; Rodighiero et al. 2007).

Consequently, from the outset the decision was taken to make the
initial selection criteria as simple and inclusive as possible, and to
then use a photometric redshift analysis to both define the final high-redshift
galaxy sample and to exclude likely low-redshift interlopers. Within this
context, the original sample for this study was a $z^{\prime}-$band selected
catalog covering the $0.63$ square-degree area uniformly covered by
the Subaru optical and UDS near-infrared data, produced with version 2.5.2
of the {\sc sextractor} software (Bertin \& Arnouts 1996). The original
catalog was then reduced to $\simeq 300,000$ objects by adopting a
$z^{\prime}-$band magnitude limit of $z^{\prime}=26$, which corresponds to the
$\simeq6\sigma$ detection threshold ($1.6''$ diameter apertures) and the $80$\% completeness limit. 
In addition, the only other selection criterion applied was the
exclusion of all objects which were detected in the $B-$band at more
than $2\sigma$ significance. This criterion was adopted in order to
exclude the vast majority of galaxies which are simply too
bright in the $B-$band to be robust high-redshift candidates; given
that the Lyman-limit at $912$\AA\, is redshifted out of the Suprime-cam
$B-$band filter at a redshift of $z=4.5$.

Following the application of the initial selection criteria the parent
sample of potential high-redshift galaxies consisted of $6495$
candidates in the apparent $z^{\prime}-$band magnitude range $23.8<z^{\prime}<26.0$.
Due to the decision not to apply strict drop-out criteria in our
initial selection, in principle, this sample should contain {\it all}
galaxies with $z^{\prime}\leq26$ lying between $z=4.5$ (when the Lyman
limit [$912$\AA] is redshifted out of the $B-$band) and $z=7.0$ (when
Lyman alpha [$1216$\AA] is redshifted out of the
$z^{\prime}$-band). However, given that the primary selection is performed in the
rest-frame UV, it is clear that the sample is still biased against
objects with significant levels of intrinsic reddening.

\subsection{Photometric redshift analysis}
The next stage in the analysis was to process each of the 
potential high-redshift candidates with our own photometric redshift
code. The photometric redshift code is an extended version of the
publicly available {\sc hyperz} package (Bolzonella, Miralles \&
Pell\'{o} 2000) and is described in more detail by Cirasuolo et
al. (2007). However, briefly, the code fits a wide range of different
galaxy SED templates to the available multi-wavelength ($BVRi'z'JK$)
photometry of each candidate, returning a best-fitting value of
redshift, SED type, age, mass and reddening. It is worth noting at
this point that for calculating the photometric redshifts, and the
final luminosity function estimates, the $1.6''$ diameter magnitudes
were corrected to total using point-source corrections in
the range $0.21$--$0.30$ magnitudes (depending on the seeing of the
$BVRi'z'JK$ images). The primary set of
stellar population models we adopted were those of Bruzual \& Charlot
(2003), with solar metallicity and assuming a Salpeter initial mass function (IMF) with a lower and upper
mass cutoff of 0.1 and  100 $M_{\odot}$ respectively. Both instantaneous burst and exponentially
declining star formation models -- with e-folding times in the range $\rm 0.1 \leq \tau (Gyr)\leq 15$ --
were included, with the restriction that the SED models did not exceed
the age of the Universe at the best-fitting redshift. The code
accounts for dust reddening by following the Calzetti et al. (2000) obscuration law
within the range $0 \leq A_V \le 6$, and accounts for Lyman series
absorption due to the HI clouds in the inter galactic medium according
to the Madau (1995) prescription. 

Based on the results of the photometric redshift analysis, 8\% of the
sample was excluded because it was not possible to find an acceptable 
$\chi^{2}$ fit (at the $3\sigma$ level) with a galaxy SED template at any redshift. The vast
majority of these objects were spurious artifacts on the
$z^{\prime}$-band images which had contaminated the original catalog. 
However $>50$\% of
the excluded contaminants with $z^{\prime}\leq24.5$ had $z^{\prime}-J$
colours consistent with ultra-cool galactic stars.

\subsection{Multiple redshift minima}
Following the photometric redshift analysis our sample consisted of
1621 high-redshift galaxy candidates with primary redshift solutions
at $z\geq4.5$, and 4350 objects with primary redshift solutions at
$z\leq4.5$. However, due to the problem of multiple redshift minima
which is inherent to photometric redshift techniques, when computing
the luminosity function is it clearly not optimal to
simply exclude the majority of candidates with primary redshift
solutions at $z\leq4.5$. Indeed, even for candidates
which have well defined primary redshift solutions at $z\geq4.5$, 
it is often the case that a plausible secondary redshift solution
exists at lower redshift; and vice-versa. The simple reason is that, within the constraints of the
available photometry, it is often statistically impossible to cleanly 
differentiate between a high-redshift Lyman-break solution and a
competing, low-redshift, $4000$\AA\,-break solution at $z\simeq1$.
Therefore, during the derivation of the luminosity functions presented
in Section 5, each candidate (including those with primary redshift
solutions at $z\leq4.5$) was represented by its normalised probability
density function $P(z)$, which satisfied the condition:
\begin{equation}
\int^{z=7}_{z=0}P(z^{\prime})\delta z^{\prime} =1
\end{equation}
\noindent
In addition to making better use of the available information, this
methodology immediately deals with the problem of multiple redshift solutions
in a natural and transparent fashion. 

\section{Luminosity function estimation}
Armed with a normalised probability density function for each
high-redshift candidate, it is possible to construct their
luminosity function using the classic $V/V_{max}$ maximum likelihood
estimator of Schmidt (1968). Therefore, within a given redshift interval
($z_{min}<z<z_{max}$), the luminosity function estimate within a given
absolute magnitude bin was calculated as follows:
\begin{equation}
\phi(M) = \sum_{i=1}^{i=N} \int^{z_{2}(m_i)}_{z_{1}(m_i)} 
\frac{C_{i}(m_i,z^{\prime})P_{i}(z^{\prime})\delta z^{\prime}}{V_{i}^{max}(m_i,z^{\prime})}
\end{equation}
\noindent
where the summation runs over the full sample, and $z_{1}\rightarrow z_{2}$ is
the redshift range within which a candidate's absolute magnitude lies
within the magnitude bin in question. The quantity $C_i(m_i,z)$ is a
correction factor which accounts for incompleteness, object blending and contamination
from objects photometrically scattered into the sample from faint-ward
of the $z^{\prime}-$band magnitude limit \footnote{A full discussion
of the influence on the estimated luminosity functions of adopting the
full photometric redshift probability
density function for each LBG candidate is provided in the appendix.}.

\subsection{Completeness and object blending}
The corrections for sample incompleteness and object blending
were calculated by adding thousands of randomly placed point-sources 
into the $z^{\prime}-$band images and recovering them with {\sc
sextractor} using the same extraction parameters adopted for the original object
catalogues. The results of this procedure revealed that the
$z^{\prime}-$band images are 100\% complete at $z^{\prime}\leq25$ and
fall to 80\% complete at the adopted $z^{\prime}=26$ magnitude limit. 
In addition, this process also demonstrated that it is necessary to
correct for a constant $\simeq15$\% of objects which are missed from the object catalogues
due to blending with nearby objects on the crowded Suprime-Cam images.

\subsection{Photometric scattering}
In this study we adopt $z^{\prime}=26$ as our limiting magnitude,
which corresponds to the $6\sigma$ detection limit. This fact,
combined with the steepness of the luminosity function in the region
we are exploring (see Fig. 1) means that the prospect of contamination 
of the faint end of our sample from objects photometrically
up-scattered from below the flux limit has to be considered carefully.
In order to quantify this effect it is necessary to rely on simulation. 
In outline, the adopted procedure was to produce a realistic simulated population of
LBGs in the redshift interval $4.0<z<6.5$, based on a
known model for the evolving luminosity function. The simulated
photometry for this population was then passed through the same initial
selection criteria and photometric redshift analysis as employed on
the real data. The average effect on the faint end of our luminosity
function determination could then be calculated by running many monte-carlo
realisations of the simulated LBG population. 

In order to make the simulated LBG population as realistic as possible
we based our model for the evolving luminosity function on the
Schechter function fits to the latest determination of the UV-selected luminosity functions centred on 
$z=3.8$, $z=5.0$ and $z=5.9$ by Bouwens et al. (2007). Over the full redshift range the adopted model
had a fixed normalisation ($\phi^{\star}=0.0015$ Mpc$^{-3}$), a fixed
faint-end slope ($\alpha=-1.75$) and an evolving characteristic
magnitude parameterised as:
\begin{equation}
M_{1500}^{\star}(z)=-21.0+1.35\log(z-2.8)
\end{equation}
\noindent
where $M_{1500}^{\star}$ is calculated at a rest-frame wavelength of
1500\AA\, (which corresponds to the middle of the $z^{\prime}-$filter
for an object at $z=5$). Although simple, this parameterisation nonetheless closely reproduces the Bouwens et
al. (2007) luminosity function fits over the redshift range of
interest. More importantly, this model should provide an accurate
prediction of the relative numbers of objects just above and just
below our adopted magnitude limit, the region in which our sample is
most vulnerable to potential contamination from photometric scatter.

This parameterisation of the evolving luminosity function was used to
populate the redshift$-$magnitude plane using bin sizes of $\Delta
z=0.1$ and $\Delta m=0.25$. Within each redshift bin the luminosity
function was integrated down to an absolute magnitude 
corresponding to $z^{\prime}=27$, one magnitude fainter than our
adopted magnitude limit. In addition to its absolute magnitude and
redshift, each simulated LBG was allocated $BVRi^{\prime}z^{\prime}JK$
photometry based on an appropriate model SED drawn randomly from a
distribution of parameter values (i.e. age, reddening, metallicity)
representative of those displayed by the real LBG sample. Appropriate
photometric errors were then calculated for the synthetic photometry
in each filter, according to the average depths of the real data.

The final stage of the process involved the use of monte-carlo
simulation whereby, in each realisation, the photometry of 
each candidate was randomly perturbed according to its photometric
errors. The perturbed photometry of the synthetic LBG sample was then 
passed through our initial optical selection criteria to establish the
number of objects which would have been included in our sample. The
final result of this process was a finely sampled grid of correction
factors which accounted for the fraction of the objects scattered
into, and out of, the sample as a function of redshift and apparent
$z^{\prime}-$band magnitude. 

In terms of the estimated luminosity functions presented in Section 5, 
this simulation process demonstrated that photometric up-scatter from 
below the magnitude limit has a relatively minor effect. As would be
anticipated, the effect is most pronounced on the two faintest absolute
magnitude bins in the $z=5$ and $z=6$ luminosity function estimates. 
However, even here the effect is only to increase the volume density
by $\simeq25$\% and $\simeq 15$\% respectively. The luminosity
function estimates shown in Fig. 1 are corrected for this effect.

\begin{figure*}
\centerline{\psfig{file=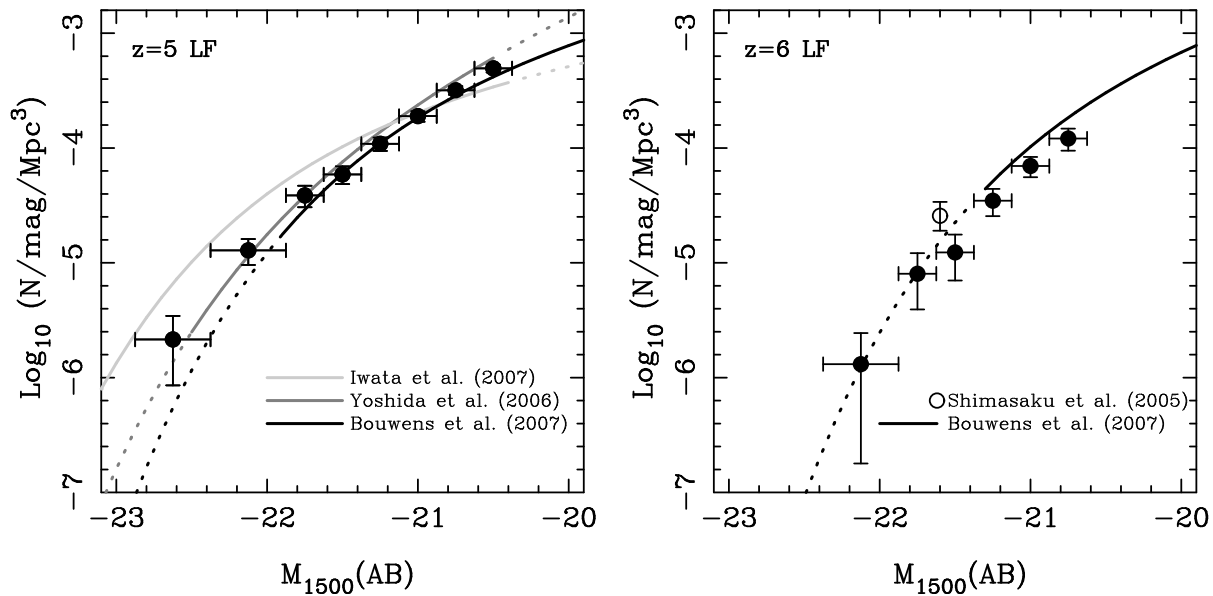,width=16.0cm,angle=0}}
\caption{A comparison of our estimates of the bright end of the
luminosity function at $z=5$ and $z=6$ with the Schechter function
fits derived by recent literature studies.
The left-hand panel compares our luminosity function estimate at $z=5$ 
with the results of Iwata et al. (2007), Yoshida et al. (2006) and Bouwens et
al. (2007). The right-hand panel compares our luminosity function
estimate at $z=6$ with the results of Shimasaku et al. (2005) and Bouwens et
al. (2007). In each panel the literature luminosity function fits are
plotted as solid curves where they are well constrained by data, and
as dotted curves where there is no data, or only data with very large uncertainties.}
\end{figure*}

\section{The luminosity function at high redshift}
Following the procedures outlined in the previous two sections, the
UV-selected luminosity functions were estimated in two redshift
intervals centred on $z=5$ and $z=6$. For the $z=5$ luminosity
function we considered the redshift range $4.7\leq z \leq5.3$ and for the
$z=6$ luminosity function we considered the redshift range
$5.7\leq z \leq6.3$. These redshift ranges were chosen to be wide enough to include sufficient
numbers of objects to provide a robust measurement of the bright end
of the luminosity function, and narrow enough to limit the evolution
of the luminosity function across each bin. The
resulting estimates of the luminosity functions at $z=5$ and $z=6$ are
shown in Fig. 1.

The advantages provided by the large area and deep
optical+nearIR data available in the UDS/SXDS are immediately 
obvious from Fig. 1. As a result of the current
data-set covering an area of 0.63 square degrees, it has been possible
to robustly estimate the bright end of the luminosity function 
to absolute magnitudes as bright as
$M_{1500}\simeq-22.5$ and $M_{1500}\simeq-22$ at $z=5$ and $z=6$ respectively.
A second obvious feature of Fig. 1 is the significant evolution in the
bright end of the luminosity function within the $5<z<6$ redshift interval; as
previously demonstrated by Bouwens et al. (2007). From the data
presented in Fig. 1 it is tempting to assume that this evolution is due
entirely to a change in $M_{1500}^{\star}$ between $z=5$ and $z=6$ (a
shift of $\Delta M_{1500}^{\star}\simeq0.5$ magnitudes results in excellent agreement
between the two luminosity functions). In fact, based on our
data-set alone, it is not possible to determine whether the apparent
evolution is due entirely to a change in $M_{1500}^{\star}$, or
whether significant evolution in the normalisation ($\phi^{\star}$) is
also occurring. However, the differential change in galaxy number density as a
function of absolute magnitude does provide some information on the form of the luminosity function evolution. For
example, while the number density of $M_{1500}\simeq-21$ galaxies only decreases by a factor of
$\simeq3$ between $z=5$ and $z=6$, the number density of
$M_{1500}\simeq-22$ galaxies decreases by a factor of $\simeq10$. In
terms of Schechter function parameters, it is not possible to
reproduce this change in luminosity function shape through evolution
of $\phi^{\star}$ alone, and immediately confirms that some evolution
of $M_{1500}^{\star}$ must be taking place. An attempt to quantify the
relative contribution of evolution in $M_{1500}^{\star}$ and
$\phi^{\star}$ is pursued in Section 5.2.

\subsection{Comparison with previous studies}
Before proceeding to explore the evolution of the luminosity function
from $z=5$ to $z=6$ further, it is worthwhile comparing our results
with those of recent studies in the literature. 

\begin{figure*}
\centerline{\psfig{file=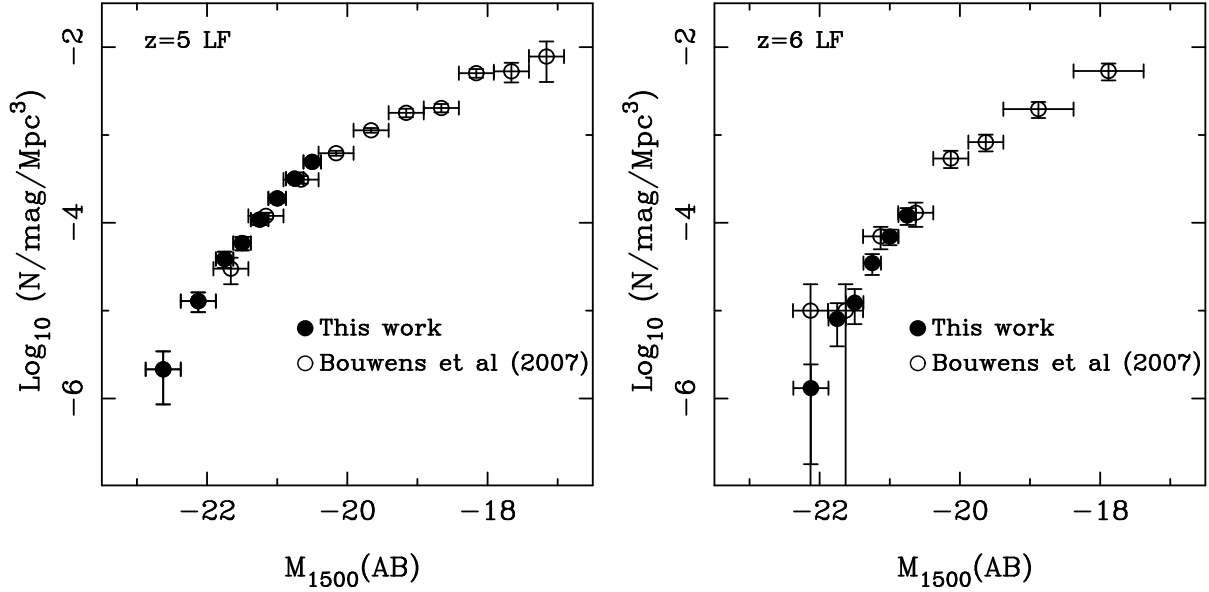,width=16.0cm,angle=0}}
\caption{A comparison of our luminosity function estimates at $z=5$ and
$z=6$ with the equivalent estimates derived from
deep HST imaging by Bouwens et al. (2007). It can be seen that the 
agreement between the two independent luminosity function estimates at
both $z=5$ and $z=6$ is excellent. This figure illustrates the
advantages provided by the wide area coverage of the current UDS/SXDS
study. Although the unparallelled sensitivity of HST imaging 
provides tight constraints on the normalisation and faint-end slope of
the $z\geq5$ luminosity function, the wide area of the current
UDS/SXDS study provides improved constraints on the bright end of the
the luminosity function, and extends the dynamic range of the data by a
factor of $\simeq3$ in luminosity.}
\end{figure*}

\subsubsection{The $z\simeq5$ luminosity function}
The $z\simeq5$ luminosity function has been investigated by several
recent studies using both ground-based and HST imaging (e.g. Ouchi et
al. 2004; Yoshida et al. 2006; Iwata et al. 2007; Oesch et al. 2007;
Bouwens et al. 2007). As an illustration of the range of results which
have recently been reported we show in Fig. 2a a comparison of our
$z=5$ luminosity function estimate with the ground-based results of
Yoshida et al. (2006) and Iwata et al. (2007), and the HST-based results of Bouwens et al. (2007).
The Yoshida et al. (2006) study is based on deep Subaru Suprime-cam optical imaging data
covering an area of $\simeq0.25$ square degrees in the Subaru Deep
Field. Yoshida et al. selected a large LBG sample centred on $z=4.7$, 
based on cuts in the $V-i^{\prime}$ vs $i^{\prime}-z^{\prime}$
colour-colour diagram, the reliability of which were confirmed via
limited follow-up spectroscopy. The Schechter function fit derived by Yoshida
et al. from their $z=4.7$ LBG sample is shown as the dark grey curve
in Fig. 2a. It can be seen that our luminosity function
estimate is in excellent agreement with the Yoshida et
al. determination at $M_{1500}\leq-21.5$. At $M_{1500}\geq-21.5$ our
number densities are somewhat lower than determined by Yoshida et al.,
although only by $\simeq20$\%. This difference is not very
significant, and can largely be explained by the fact that the
Schechter function fit derived by Yoshida et al. does not have a well
constrained faint-end slope due to the lack of data fainter than
$\simeq M_{1500}^{\star}$. We note here that the Yoshida et al. results at $z\simeq5$ are
very similar to those of Ouchi et al. (2004a), who analysed an
combined SDF+SXDS data-set covering a larger area (0.35 square
degrees) but with a magnitude limit $\simeq0.5$ magnitudes brighter.

The solid black curve plotted in Fig. 2a is the Schechter function fit derived
by Bouwens et al. (2007) from a sample of 1416 $V-$drop galaxies
selected from a compilation of several deep HST imaging surveys
(GOODS N+S, HUDF \& HUDF-P). At $M_{1500}\geq-22$ our data are in
excellent agreement with the Bouwens et al. luminosity function, which
is remarkable given the very different techniques, data-sets and sky
areas utilised by the two studies (the largest area in the Bouwens et
al. study is the combined GOODS N+S fields; $\ltsim 0.1$ square
degrees). At the brightest magnitudes Fig. 2a suggests that our sample
contains a higher number density of galaxies than predicted by the Bouwens
et al. luminosity function. However, this is not unreasonable given that the large survey area of
this study provides improved statistics on the bright end of the luminosity
function, where the Bouwens et al. fit to the luminosity function is less well constrained. 

Finally, in Fig. 2a
we plot the recent luminosity function fit of Iwata et al. (2007)
which is based on deep Subaru Suprime-cam imaging covering an area of
$\simeq0.35$ square-degrees (HDF-N and the J0053+1234 region). Based on cuts in the $V-I_C$ vs $I_C-z^{\prime}$
colour-colour diagram, Iwata et al. (2007) select 853 $z\simeq5$ LBG candidates
and derive the luminosity function fit shown as the light grey curve in Fig. 2a.
It can be seen that our luminosity function estimate is in poor
agreement with the Iwata et al. (2007) results, with the latter study
finding a number density of objects at the
brightest absolute magnitudes a factor of $\simeq3$ higher than found
here. It is difficult to understand how the
large numbers of bright LBGs detected in the Iwata et al. study could
be missing from our sample. Objects this bright would have very high
signal-to-noise photometry in our data-set, and should return the most
robust photometric redshifts. The Iwata et al. (2007) results are in
good agreement with their previous estimate of the $z\simeq5$
luminosity function (Iwata et al. 2003) and it has been suggested by
several authors (e.g. Ouchi et al. 2004a; Yoshida et al. 2006; Bouwens et al. 2007) that
the selection criteria used in Iwata et al. (2003, 2007) may allow
contamination from low-redshift interlopers. We note that to
produce number densities at $M_{1500}\leq-22$ as high as reported by
Iwata et al. (2007), we would have to reinstate into our sample objects which have been
excluded as suspected ultra-cool galactic stars due to 
their photometry being a poor match to any galaxy SED template

\begin{figure*}
\centerline{\psfig{file=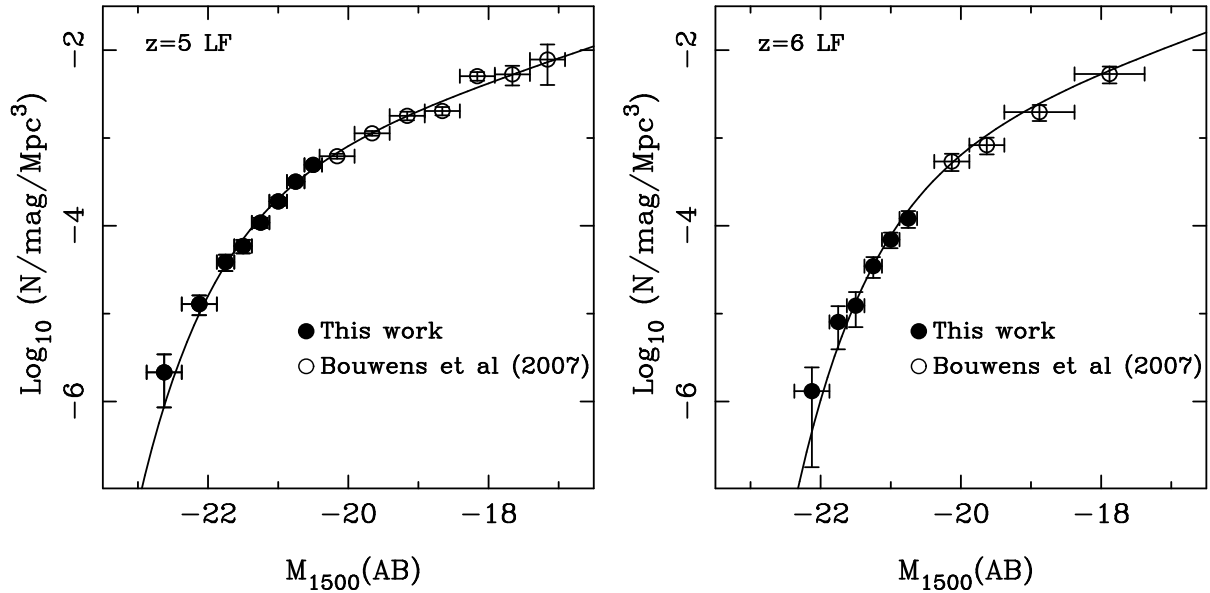,width=16.0cm,angle=0}}
\caption{The best-fitting Schechter functions to the combined
(ground-based+HST) data-sets at $z=5$ and $z=6$ are shown as the solid
curves in the left-hand and right-hand panels respectively. During the
fitting process the bright end of the luminosity function is
constrained using the data from the current study (filled circles),
while the faint end is constrained using the results of Bouwens et
al. (2007), which are based on small area, but ultra-deep HST
imaging. The best-fitting Schechter function parameters are listed in
Table 1.}
\end{figure*}

\subsubsection{The $z\simeq6$ luminosity function}

Previous constraints on the $z\simeq6$ luminosity function have
been based, virtually exclusively, on the deep HST imaging available in
the two GOODS regions, the Hubble Ultra-Deep Field (HUDF) and the HUDF
parallel fields (e.g. Bunker et al. 2004; Dickinson et al. 2004; Yan
\& Windhorst 2004; Malhotra et al. 2005; Bouwens et al. 2006,
2007). In Fig. 2b we show the Schechter function fit to the $z\simeq6$
     luminosity function derived by Bouwens et al. (2007) from a
     sample of 627 $i-$dropouts selected from a compilation of the
     deepest available HST imaging data\footnote{the reader is also
     referred to Bouwens et al. (2007) for a detailed comparison of
     previous constraints on the $z\simeq6$ luminosity function based
     on HST imaging data}. Fig. 2b shows that our estimate of the
     $z=6$ luminosity function is in reasonable agreement with the
     luminosity function fit of Bouwens et al. (2007), although our number densities
     are lower by $\simeq40$\% in the region around
     $M_{1500}\simeq-21$, where the Bouwens et al. luminosity
     function fit is still well constrained by data. However, this
     discrepancy is only apparent in comparison to the
     Schechter function fit derived by Bouwens et al., and the actual
     data from the two studies are in excellent agreement (see
     Fig. 3b). In Fig. 2b we also show the number density of objects
     at $M_{1500}=-21.6$ derived by Shimasaku et al. (2005), based on a
     sample of 12 objects selected using imaging of the Subaru Deep
     Field with two medium-band $z-$filters. The Shimasaku et
     al. estimate is a factor of $\simeq2$ higher than our estimated
     number density at $M_{1500}=-21.6$. The source of this
     discrepancy is not clear. Given the rigorous nature of
     their selection criteria, it seems unlikely that the Shimasaku et
     al. sample is contaminated by interlopers at the $\simeq50$\%
     level. However, the survey area covered by this work is a factor
     of $\simeq3$ larger than that of the Shimasaku study, so it is
     likely that the apparent off-set is simply due to a combination
     of statistical errors and cosmic variance.

\subsection{Combining the ground-based and HST data-sets}
Although the results of the present study provide much improved
constraints on the bright end of the luminosity function
at $z=6$ and $z=5$, they do not reach faint enough to constrain the
normalisation or faint-end slope of the $z\geq5$ luminosity
function. Consequently, in this section we explore the possibility of
providing an improved measurement of the evolution of the luminosity
function between $z=5$ and $z=6$ by combining our results with those
of the HST-based study of Bouwens et al. (2007). In Fig. 3 we show a
comparison of our luminosity function estimates at $z=5$ and $z=6$ with those of Bouwens
et al. (2007). It can be seen that in the absolute magnitude range
common to both studies the two, independent, estimates of the luminosity functions are in
excellent agreement. Motivated by this agreement, it was decided to
combine our results with those of Bouwens
et al. (2007) to derive Schechter function fits to the luminosity
function at $z=5$ and $z=6$ based on data spanning a factor of
$\simeq100$ in luminosity.

\subsubsection{Fitting the luminosity functions}
The adopted procedure was to use a maximum likelihood technique to fit
a Schechter function to a finely sampled grid on the
apparent $z^{\prime}-$band magnitude $-$ redshift plane $(m_{z}-z)$. To fit the
luminosity function at $z=5$ the grid was populated within the redshift
interval $4.7<z<5.3$ and the apparent $z^{\prime}-$band magnitude range
$23<m_{z}<29$. In the magnitude range $23<m_{z}<26$ the grid was
populated using the data from this study, while in the magnitude range 
$26<m_{z}<29$ the grid was populated according to the Schechter
function fit to the $z=5$ luminosity function derived by Bouwens et
al. (2007). This process was repeated to fit the $z=6$ luminosity
function, with the exception that the redshift range considered was $5.7<z<6.3$.
The fitting procedure was to maximise the following likelihood function:
\begin{equation}
\log L = \Sigma_{i,j} n_{i,j}\log p_{i,j}
\end{equation}
\noindent
where $n_{i,j}$ is the number of galaxies in cell $(i,j)$ and $p(i,j)$
is the probability of finding a galaxy within cell $(i,j)$ given the
choice of model parameters. In this context, the probability $p(i,j)$ is naturally
defined as $n_{p}(i,j)/\Phi_{z}$, where $n_{p}(i,j)$ is
the predicted number of galaxies within cell $(i,j)$, for a given set of
model parameters, and $\Phi_{z}$ is the corresponding integrated
luminosity function.

\subsubsection{Results}
The best-fitting Schechter functions to the combined ground-based+HST 
data-sets at $z=5$ and $z=6$ are shown as the solid curves in
Fig. 4. The best-fitting Schechter function parameters and their associated
uncertainties are reported in Table 1, and the $1\sigma, 2\sigma$
and $3\sigma$ confidence regions around the best-fitting values of
the faint-end slope and characteristic magnitude are plotted in~Fig.~5.

It can be seen from Fig. 4 that a Schechter function provides a good
description of luminosity function at $z=5$ and $z=6$ over the full
luminosity range ($\geq5$ magnitudes) sampled by the combined
ground-based+HST data-set. A comparison of the best-fitting Schechter
function parameters derived here (Table 1) with those reported by
Bouwens et al. (2007) shows that, at both redshifts, our best-fitting
parameters are within the $1\sigma$ uncertainties reported by the
earlier study. With respect to the faint-end slope and
normalisation this is unsurprising, given that the constraints on
these two parameters are largely provided by the deeper HST imaging data.
However, the principle advantage provided by fitting the combined
data-set is the improved constraints on the value of
$M^{\star}_{1500}$; particularly at $z=6$. Our results suggest that 
$M^{\star}_{1500}$ dims by $0.69\pm0.16$ magnitudes between $z=5$ and
$z=6$. This is more dramatic than, although still completely consistent with, the
dimming of $0.40\pm0.23$ suggested by the Schechter function fits
derived by Bouwens et al. (2007). With respect to the normalisation of
the luminosity function, our results suggest that $\phi^{\star}$ increases
by a factor of $\simeq2$ between $z=5$ and $z=6$, although the
comparatively large uncertainty on the normalisation of the Schechter
function at $z=6$ means that this evolution is not statistically
significant ($<2 \sigma)$. As expected, our fit to the combined
ground-based+HST data-set confirms the Bouwens et al. (2007) result
that the faint-end slope of the luminosity function at $z\geq5$ is
steep ($\alpha\simeq -1.7$) and shows no sign of evolution in the
redshift interval $5<z<6$. 

\begin{figure}
\centerline{\psfig{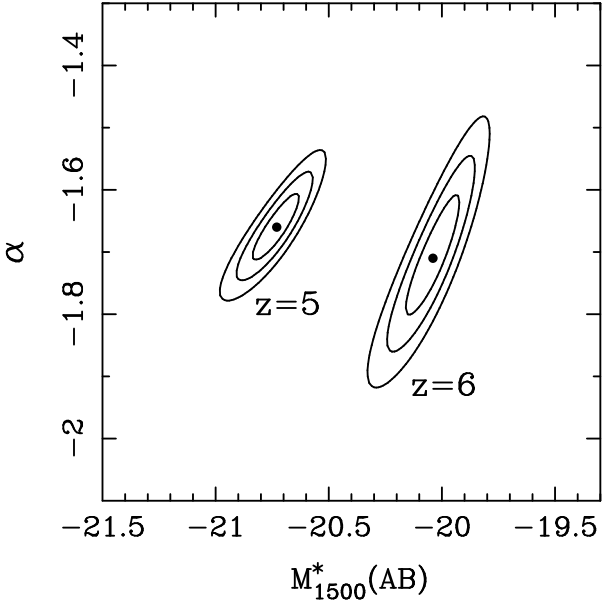}}
\caption{Constraints on the faint-end slope and characteristic
magnitude of the luminosity function at $z=5$ and $z=6$ derived via 
maximum likelihood fitting of the combined (ground-based+HST) data-set.
The location of the best-fitting parameters at each redshift are shown
by filled circles, and the plotted contours indicate the $1\sigma, 2\sigma$
and $3\sigma$ confidence regions.}
\end{figure}
\begin{table}
\caption{The best-fitting Schechter function parameters from our
maximum likelihood fitting of the combined (ground-based+HST)
luminosity function data-sets at $z=5$ and $z=6$. The errors on the
faint-end slope and characteristic magnitude have been derived from
the likelihood ratio contours shown in Fig. 5. The values of
$\phi^{\star}$ and their associated errors have been calculated by
fitting the number counts in each redshift interval.}
\begin{tabular}{cccc}
\hline
Redshift & $\phi^{\star}$/Mpc$^{-3}$ & M$^{\star}_{1500}$ & $\alpha$\\
\hline
5.0      & $(9.4\pm{1.9})\times 10^{-4}$ & $-20.73\pm{0.11}$ & $-1.66\pm{0.06}$\\
6.0      & $(1.8\pm{0.5})\times 10^{-3}$ & $-20.04\pm{0.12}$ & $-1.71\pm{0.11}$\\
\hline
\end{tabular}
\end{table}

\section{Clustering analysis}
In this section we evaluate the 2-point angular correlation function
for those LBGs in our sample with primary photometric redshift
solutions in the interval $5.0<z<6.0$. To measure the angular correlation function
$\omega(\theta)$ and estimate the related poissonian errors we used
the Landy \& Szalay (1993) estimators. The correlation
function derived from our $5.0<z<6.0$ LBG sample is shown in Fig. 6.  
The best fit for the angular correlation is assumed to be a power-law
as in Groth \& Peebles (1977):
\begin{equation}
\omega(\theta)=A_{\omega}(\theta^{-\delta}-C_{\delta})
\end{equation}
\noindent
with the amplitude at 1 degree $A_{\omega}=13.0^{+5.9}_{-4.1}\times10^{-3}$,
the slope $\delta=0.72\pm0.08$, and the integral constraint due to the
limited area of the survey $C_{\delta}=2.34$ (determined over the
unmasked area). It should be noted that because the contamination of
our $5<z<6$ LBG sample from low-redshift interlopers is securely $<10\%$ we have
made no correction for dilution of the clustering signal. 

\begin{figure}
\leftline{\psfig{file=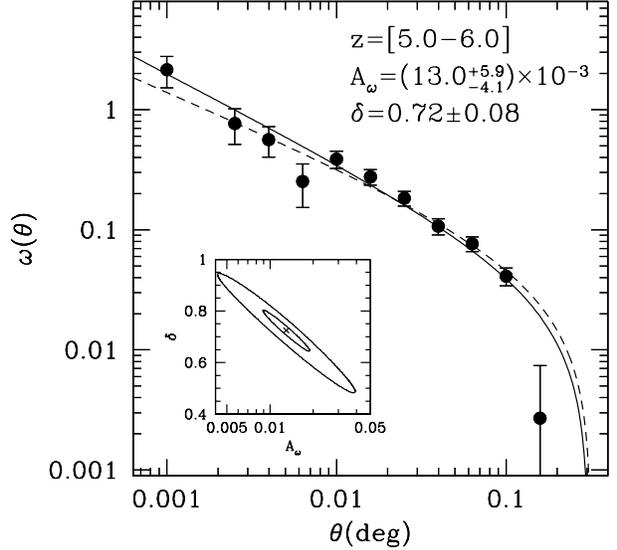,width=9.0cm,angle=0}}
\caption{2-point angular correlation function determined for our
     sample of $5.0<z<6.0$ LBGs, with its fitted power-law
     (solid line). The best-fitting parameters for the amplitude and
     slope are indicated. The dashed line represents the power-law
     fitting only the largest scales measurements ($\theta>10''$)
     assuming a slope fixed to $\delta=0.6$. The plot inset presents
     the $\chi^2$ minimisation to estimate the best-fitting values of
     $A_{\omega}$ and $\delta$, with contours showing the $1\sigma$ and
     $3\sigma$ confidence levels.}
\end{figure}

\begin{figure*}
\centerline{\psfig{file=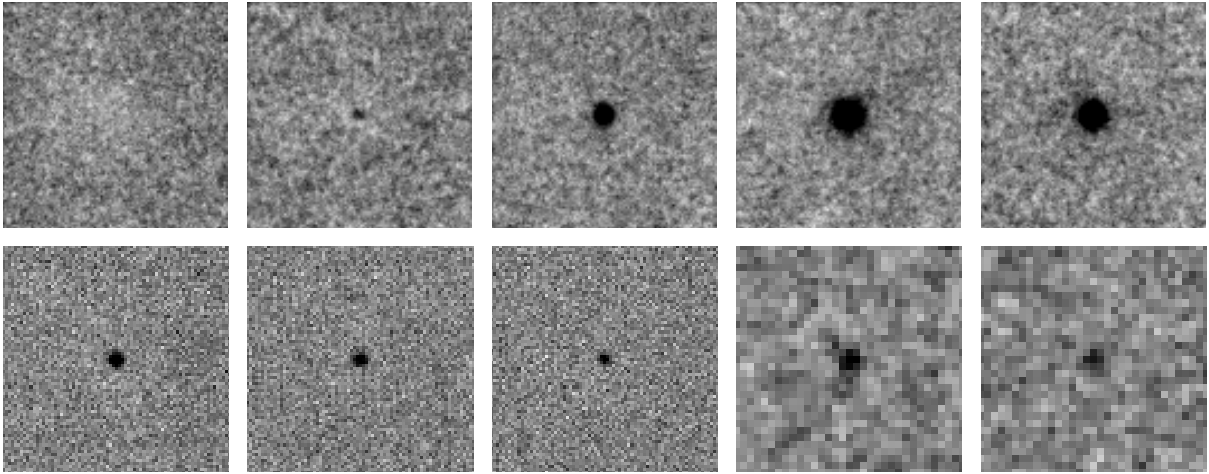,width=16.0cm,angle=0}}
\caption{Postage stamps of the stacked imaging data for the 754 LBGs
     with primary photometric redshift solutions in the range
$5.0<z<6.0$. From top left to bottom right the first eight postage
stamps show the stacked $BVRi^{\prime}z^{\prime}JHK$ data. The
final two postage stamps show the stacked IRAC data at 3.6$\mu$m
and 4.5$\mu$m. It can be seen from the second panel that the stacking
analysis produces a very faint detection in the $V-$band
($V=32.3\pm0.6$). This is not unexpected given that the Lyman-limit
does not pass fully through the Subaru $V-$band filter until $z=5.5$.
All the postage stamps have dimensions of
20\asec$\times$ 20\asec, and are shown on a linear grey-scale in the range
$\pm 5\sigma$ (where $\sigma$ is the rms sky noise). The $H-$band
photometry recently became available in UKIDSS DR3, and was only used
in the stacking analysis.}
\end{figure*}

In Fig. 6 we also indicate the best-fitting curve
determined using only the largest scale measurements ($\theta>10''$)
and a fixed value for the slope of $\delta=0.6$. Based on a sample of
$z\simeq5$ LBGs selected from the GOODS fields, Lee et al. (2006) find
that their angular clustering measurments show a strong
transition on the smallest scales ($\theta<10''$), indicative of a intra/inter dark
matter halo transition, while on larger scales the angular correlation
is well approximated by a power-law with a slope of
$\delta=0.6$. However, as can be seen from Fig. 6, our results 
are consistent with a value of $\delta=0.6$ on large scales, and only
present a non-significant deviation from this slope on the
smallest scales ($\theta=3.6''$ corresponding to $\sim 20 h^{-1}_{70}$kpc).

In order to compare the clustering of galaxy populations at different
redshifts it is necessary to derive spatial correlation
measurements. Using the redshift distribution of our sample we can derive the
correlation length using the relativistic Limber equation
(Magliocchetti \& Maddox 1999) and, in addition, derive a linear bias
estimation (Magliocchetti et al. 2000) assuming that the dark matter
behaves as predicted by linear theory. To perform this calculation we
have adopted the redshift distribution for our sample as derived from 
integrating over the redshift
probability density functions for those objects with primary
photometric redshift solutions in the interval $5.0<z<6.0$. The correlation length
can then be evaluated by fixing the slope of the correlation function to $\gamma=1+\delta=1.72$. 
This calculation produces a measurement of $r_0=8.1^{+2.1}_{-1.5}\,h^{-1}_{70}$Mpc
for the correlation length and a value of $b=5.4^{+1.2}_{-0.8}$ for
the linear bias. Although we consider this to be our best estimate of
the clustering properties of the sample, we note that if we instead
adopt the redshift distribution based on the best-fitting photometric
redshifts alone ($\bar{z}=5.38$; see Fig A2) we derive slightly higher, although consistent,
values of $r_0=9.6^{+2.5}_{-1.8}\,h^{-1}_{70}$Mpc and $b=6.2^{+1.4}_{-1.0}$
for the correlation length and linear bias respectively.

It is of interest to compare our clustering results for
$z^{\prime}<26.0$ LBGs in the redshift interval $5.0<z<6.0$ with
recent literature results for LBG samples with similar limiting
magnitudes and mean redshifts. Our results are in good agreement with 
Ouchi et al. (2004b) who derived a correlation length of
$r_0=8.4^{+1.8}_{-2.4}\,h^{-1}_{70}$Mpc for a sample 
of $z^{\prime}<25.8$ LBGs with a mean redshift of $\bar{z}=4.7$ in the Subaru Deep Field. 
In addition, our results are in agreement with those of
Overzier et al. (2006) who derived a correlation length of
$r_0=9.6^{+4.0}_{-5.6}\,h^{-1}_{72}$Mpc for a sample of
$z_{850}<27.0$\footnote{$z_{850}$ refers to $z-$band magnitudes
obtained through the HST F850LP filter, rather than the Subaru $z^{\prime}-$filter.}
LBGs with a mean redshift of $\bar{z}=5.9$ in the GOODS
fields. Finally, our results are also in good agreement with those of Lee et al. (2006), who
derived a value of $r_0=7.5^{+1.1}_{-1.0}\,h^{-1}_{70}$Mpc (fixed
slope of $\delta=0.6$) from their sample of $z_{850}<26.0$ LBGs in the 
GOODS fields with a mean redshift of $\bar{z}=4.9$. 

Based on the predictions of dark matter halo models (Sheth et
al. 2001; Mo \& White 2002), and assuming a one-to-one correspondance between galaxies and
halos, we infer that our $5.0<z<6.0$ LBGs are likely to be hosted by 
dark matter halos with masses of $M_{DM}\geq10^{11.5-12}M_{\odot}$,
comparable to what is observed for $L>L^{\star}$ LBGs at $z\sim3-4$. 

\section{Stacking analysis}
In this section we explore the possibility of estimating the typical stellar mass
of our $5<z<6$ LBGs from a stacking analysis of
the available optical+nearIR imaging data. The aim is to test whether 
the typical stellar mass is compatible with the dark matter halo mass 
suggested by the clustering analysis.

In Fig. 7 we show 20\asec$\times$ 20\asec\, postage stamps produced by
taking a median stack of the available imaging data, centred on the
positions of the 754 LBG candidates with primary photometric redshift
solutions in the interval $5.0<z<6.0$. There are several features of 
Fig. 7 which are worthy of comment. Firstly, in the stack of the $B-$band
photometry the LBG candidates are robustly non-detected ($B\leq33$;
$1\sigma$), as expected from our initial selection criteria. Secondly,
it can be seen that there is a faint detection in the stacked $V-$band
data. However, the $V-$band detection is very faint ($V=32.3\pm0.6$)
and is expected given that the Lyman-limit does not pass out of the
Subaru Suprime-Cam $V-$filter until $z=5.5$. Finally, the two panels
on the lower right of Fig. 7 show the results of stacking the
available Spitzer IRAC data at $3.6\mu$m and $4.5\mu$m from the SWIRE
survey (Lonsdale et al. 2003). Although relatively shallow, the SWIRE
data is deep enough to produce detections at $3.6\mu$m and $4.5\mu$m
in the median stack. Together with the robust photometry from the
stacked $JHK$ imaging data, it is the extra wavelength coverage
provided by the detections at $3.6\mu$m and $4.5\mu$m which will allow
us to place useful constraints on the typical stellar mass of the
$5<z<6$ LBGs.

\subsection{SED fitting}
Using the same procedure as described in Section 3.1, the best-fitting
galaxy template to the stacked photometry is shown in
Fig. 8. It can be seen from the lower panel of Fig. 8 that the fit to
the stacked photometry provides a robust redshift solution at
$z=5.43^{+0.05}_{-0.10}$ and that any alternative low-redshift solution is securely ruled
out. The best-fitting SED template has an age of 400 Myrs, consistent
with a typical formation redshift of $z\simeq7-8$, and no intrinsic reddening ($A_{V}=0$). 
By considering the range of SED templates which provide an acceptable
fit to the stacked photometry, we calculate that the stellar mass of
the LBG stack is $\log_{10}({\rm{M}/\Msolar})=10.0^{+0.2}_{-0.4}$
(Salpeter IMF). 

\subsection{Dark matter to stellar mass ratio}
Although the uncertainties in the determination of the typical stellar and dark
halo mass for the $5<z<6$ LBG sample are obviously large, combined
with the results of the clustering analysis reported in the previous
section, the SED fit to the stacked photometry suggests that the
typical dark matter to stellar mass ratio for the $5<z<6$ LBGs is 
$\frac{M_{DM}}{M_{stars}}\simeq 30-100$.
Given that our LBG sample are exclusively drawn from the bright end of
the $5<z<6$ luminosity function it would seem reasonable to assume that they will evolve into objects
occupying the high-mass end of the low-redshift galaxy stellar mass
function (i.e. $M\geq M^{\star}_{stars}$). In which case, it is 
noteworthy that $\frac{M_{DM}}{M_{stars}}\simeq50$ is in good
agreement with the measured value for Luminous Red Galaxies
(LRGs) in the SDSS from galaxy-galaxy lensing (Mandelbaum et
al. 2006). Moreover, $\frac{M_{DM}}{M_{stars}}\simeq50$ is also 
in good agreement with the predicted mean value for low-redshift LRGs
from recent semi-analytic galaxy formation models (e.g. Almeida et al. 2008)

\begin{figure}
\leftline{\psfig{file=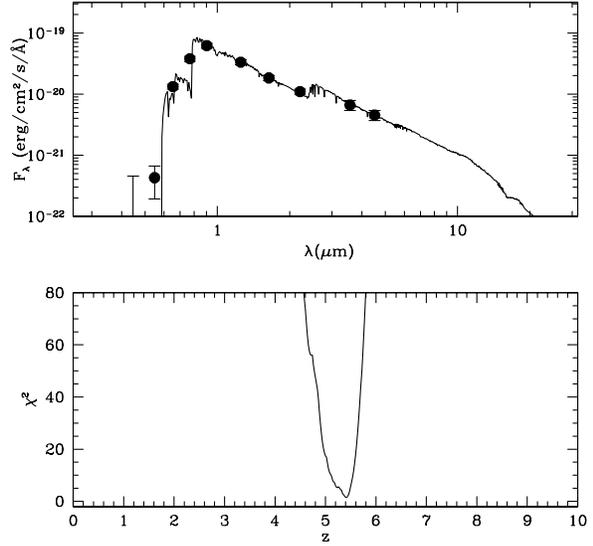,width=8.0cm,angle=0}}
\caption{The upper panel shows the SED fit to the stacked imaging 
data of those LBG candidates with primary photometric redshift solutions in
the range $5.0<z<6.0$.  In addition to the $BVRi^{\prime}z^{\prime}JK$
photometry used in the selection of the LBG sample, the stacked
data includes $H-$band (which recently became available in
UKIDSS DR3), and detections at 3.6$\mu$m and 4.5$\mu$m from stacking
the Spitzer SWIRE data covering the UDS. The bottom panel shows $\chi^{2}$
versus redshift for the SED fit to the stacked data. As
expected, the SED fit has a very robust solution at $z=5.43^{+0.05}_{-0.10}$ and
no plausible low-redshift solutions are evident.}
\end{figure}

\section{The stellar mass function}
Although it remains a possibility that a substantial population of
massive, comparatively red, objects exist at high redshift, it seems likely
that the luminous ($L\geq L^{\star}$) LBGs selected in this study are
amongst the most massive galaxies in existence at $5<z<6$. As such,
their number densities and stellar masses have the potential to
provide constraints on the latest generation of galaxy evolution models. 

In Fig. 9 we show an estimate of the galaxy stellar mass function at
$5<z<6$ based on combining the luminosity function fits presented in
Section 5 with the results of the stacking analysis presented in Section 7. 
The adopted procedure for estimating the stellar mass function is
extremely straightforward. Based on the Schechter function fits to the
luminosity function at $z=5$ and $z=6$, we have
estimated the corresponding stellar mass function by simply
multiplying the luminosity function fits by the mass-to-light ratio derived from our SED fit to the
stacked LBG imaging data. In Fig. 9 the upper envelope 
of the grey shaded area is our estimate of the stellar mass function
at $z=5$, while the lower envelope is our corresponding estimate at
$z=6$ (both include the $1\sigma$ uncertainties on the best-fitting
Schechter luminosity function parameters). 
For comparison we also plot the stellar mass function estimates at
$z=5.3$ based on the De Lucia \& Blaizot (2007) and Bower et al. (2006) semi-analytic galaxy
formation models (thick and thin curves respectively).\footnote{These
models are publicly available from the following website:
http://www.g-vo.org/Millennium. The model predictions at $z=5.3$ are
adopted because they are the closest available to the mean photometric
redshift of our $5<z<6$ LBG sample ($\bar{z}=5.38$).} In order to
perform a fair comparison, we have shifted our stellar mass function
estimates to lower masses by a factor of 1.8, to account for the fact that the
De Lucia \& Blaizot (2007) and Bower et al. (2006) models are based on IMFs which
typically return stellar masses a factor of $\simeq1.8$ lower than our
Salpeter-based estimates (De Lucia \& Blaizot 2007). Finally, for
reference, the dashed curve in Fig. 9 is the stellar mass function at $z\simeq0$ from Cole et al. (2001).

The mass function predictions from the Bower et al. (2006) and De Lucia \& Blaizot
(2007) models are in good agreement, at least qualitatively, with our
observational estimate. The general agreement between our estimate, based on
the UV-selected luminosity function, and the model predictions
suggests that LBGs constitute the majority of the stellar mass density
at $z\geq5$ and, if the model predictions are correct, a putative
population of reddened/older galaxies at $5<z<6$ does not dominate the
stellar mass density. Compared to the $z\simeq0$ mass function, our
estimate of the high-redshift stellar mass function suggests that the
number density of $\simeq10^{10}\Msolar$ and $\simeq10^{11}\Msolar$
galaxies in place by $z\simeq5$ is $\simeq10$\% and $\simeq 0.1$\% of its local value respectively.

\subsection{Stellar mass density}
Given that we have estimated the stellar mass function at
$z=5$ and $z=6$ by simply applying a constant mass-to-light ratio derived from our
stacking analysis, it is obviously somewhat speculative to proceed to
estimate the co-moving stellar mass density. 

However, for completeness, integrating our stellar mass function estimates down to
a mass limit of $M\geq10^{9.5}\Msolar$ \footnote{equivalent to integrating the $z=5$ and $z=6$
luminosity functions to a limit of M$_{1500}=-19.3$, which involves an extrapolation $\simeq1$ magnitude
fainter than the $z^{\prime}-$band magnitude limit of the UDS/SXDS data
set.} provides stellar mass density estimates of
$\simeq1\times 10^{7}\Msolar$ Mpc$^{-3}$ and $\simeq4\times
10^{6}\Msolar$ Mpc$^{-3}$ at $z=5$ and $z=6$ respectively (Salpeter IMF). 
It was decided to integrate our stellar mass function down to a
limit of $M\geq10^{9.5}\Msolar$ for two reasons. Firstly, a mass limit
of $M\geq10^{9.5}\Msolar$ corresponds well to the mass limits of 
Stark et al. (2007) and Yan et al. (2006), who previously estimated the
stellar mass density at $z=5$ and $z=6$ respectively. Secondly, given
that our data-set provides no information on the mass-to-light ratios
of objects with masses smaller than $10^{9.5}\Msolar$, it would clearly be 
ill advised to extrapolate to smaller stellar masses.

Our stellar mass density estimate at $z=5$ is in reasonable agreement with the
estimate of $6\times 10^{6}\Msolar$ Mpc$^{-3}$ derived by Stark et
al. (2007) from a combined sample of spectroscopic and photometrically 
selected galaxies in the southern GOODS field. Moreover, our
estimate of the stellar mass density at $z=6$ is in
agreement with the lower limit of $1.1-6.7\times 10^{6}\Msolar$
Mpc$^{-3}$ derived by Yan et al. (2006), based on a sample of $i-$drop
galaxies in the north and south GOODS fields. The stellar mass
densities quoted by both Stark et al. (2007) and Yan et al. (2006) are
also based on a Salpeter IMF.

\begin{figure}
\centerline{\psfig{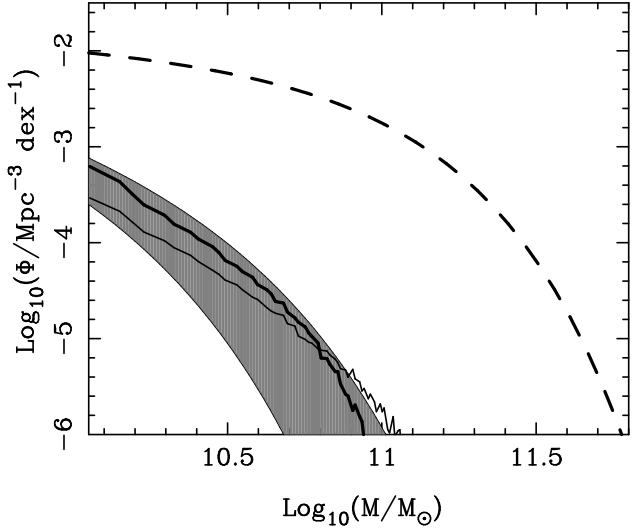}}
\caption{Estimates of the stellar mass function in the redshift
interval $5<z<6$. The upper envelope of the grey shaded area is 
our estimate of the stellar mass function at $z=5$, and 
the lower envelope is our corresponding estimate at $z=6$ (both
include the $1\sigma$ uncertainties on the best-fitting Schechter
luminosity function parameters). The thick
and thin solid lines show the estimated stellar mass function at
$z=5.3$ derived from a combination of the Millennium N-body simulation
and the semi-analytic galaxy formation models of De Lucia et
al. \& Blaizot (2007) and Bower et al. (2006) respectively. The upper dashed
curve is the stellar mass function measurement at $z\simeq0$ from Cole et al. (2001).
}
\end{figure}

\section{Conclusions}
In this paper we have reported the results of a study of a large 
sample of luminous ($L\geq L^{\star}$) LBGs in the redshift interval
$4.7<z<6.3$. By employing a photometric redshift analysis of the available
optical+nearIR data we have derived improved estimates of the
bright end of the UV-selected luminosity function at $z=5$ and $z=6$. 
Moreover, by combining our new results with those
based on deeper, but small area, HST data we have derived improved
constraints on the best-fitting Schechter function parameters at $z=5$ and $z=6$.
In addition, by studying the angular clustering properties of our
sample we have determined that luminous LBGs at $5<z<6$ typically lie in dark matter halos with
masses of $10^{11.5-12}\Msolar$. Finally, based on the results of a
stacking analysis, we have estimated the galaxy stellar mass functions and
integrated stellar mass densities at $z=5$ and $z=6$. Our main
conclusions can be summarised as follows:

\begin{enumerate}

\item{Our new determination of the bright end of the high-redshift 
luminosity function confirms that significant evolution occurs over
the redshift interval $5<z<6$. Based on our results it is clear
that the luminosity function evolution cannot be described by
evolution in normalisation ($\phi^{\star}$) alone, and that some level
of evolution in $M^{\star}_{1500}$ is also required.}

\item{A comparison of our new results with those in the literature
demonstrates that, within the magnitude range where the two
studies overlap, our estimates of the luminosity function at 
$z=5$ and $z=6$ are in excellent agreement with those derived from
ultra-deep HST imaging data by Bouwens et al. (2007).}

\item{By combining our estimate of the bright end of the luminosity
function with the corresponding estimates of the faint end by Bouwens
et al. (2007), it is possible to fit the luminosity function at $z=5$
and $z=6$ over a luminosity range spanning a factor of
$\simeq100$. Based on this combined ground-based+HST data-set we find
the following best-fitting Schechter function parameters:
$M_{1500}^{\star}=-20.73\pm0.11, \phi^{\star}=0.0009\pm0.0002\, {\rm
Mpc}^{-3}$ and $\alpha=-1.66\pm0.06$ for the luminosity function at $z=5$, and
$M_{1500}^{\star}=-20.04\pm0.12, \phi^{\star}=0.0018\pm0.0005\,{\rm
Mpc}^{-3}$ and $\alpha=-1.71\pm0.11$ at $z=6$.}

\item{These results are consistent with the corresponding Schechter
function parameters derived by Bouwens et al. (2007) although, due to
the improved statistics at the bright end provided by our wide survey
area, the fits to the combined ground-based+HST data-set provide
improved constraints on the evolution of $M^{\star}_{1500}$ in particular.}

\item{An analysis of their angular clustering properties demonstrates
that luminous $L\geq L^{\star}$ LBGs at $5<z<6$ are strongly
clustered, with a correlation length of $r_0=8.1^{+2.1}_{-1.5}\,h^{-1}_{70}$Mpc.
Comparison of these clustering results with theoretical models suggests that these
LBGs typically reside in dark matter halos with masses of $\simeq10^{11.5-12}\Msolar$.}

\item{An SED fit to a stack of the available optical+NearIR data for our sample
suggests that luminous LBGs at $5<z<6$ have typical stellar masses of
$\simeq10^{10}\Msolar$. Combined with the results of the clustering
analysis this suggests that the typical ratio of dark matter to stellar mass 
for luminous LBGs at $5<z<6$ is $\frac{M_{DM}}{M_{stars}}\simeq30-100$.}

\item{Assuming that the mass-to-light ratio derived from the SED fit
to the stacked LBG imaging data is representative, we use
our best-fitting Schechter function parameters for the $z=5$ and $z=6$
luminosity functions to estimate the corresponding stellar mass functions.
Although clearly subject to large uncertainties, our stellar mass
function estimates are consistent with the latest predictions of the semi-analytic galaxy
formation models of De Lucia \& Blaizot (2007) and Bower et al. (2006).}

\item{Based on our stellar mass function estimates we calculate that
the stellar mass in place at $z=5$ and $z=6$ is $\simeq1\times
10^{7}\Msolar$ Mpc$^{-3}$ and $\simeq4\times 10^{6}\Msolar$ Mpc$^{-3}$
respectively. Although uncertain, these independent estimates of the
integrated stellar mass density are consistent with the results of
studies utilising the ultra-deep Spitzer IRAC data in the GOODS 
fields (Stark et al. 2007; Yan et al. 2006).}

\end{enumerate}

\section{acknowledgments}
The authors would like to thank the anonymous referee whose comments
and suggestions significantly improved the final version of this
manuscript. RJM and OA would like to acknowledge the funding of the Royal
Society. MC and SF would like to acknowledge funding from STFC.
We are grateful to the staff at UKIRT and Subaru for making these 
observations possible. We also acknowledge the Cambridge Astronomical
Survey Unit and the Wide Field Astronomy Unit in Edinburgh for processing the UKIDSS data.

\begin{appendix}

\section{Luminosity function estimates: technical details}
When deriving our estimates of the LBG luminosity functions at $z=5$
and $z=6$ in Section 4 we employed several techniques which differ
somewhat from those which have been commonly adopted in previous literature
studies. Consequently, it is perhaps instructive to investigate what
influence these techniques have on the derived luminosity function
estimates. In this section we first investigate the influence of our
completeness corrections, before proceeding to investigate the
influence of our decision to make use of the full photometric redshift
probability density function for each LBG candidate.

\subsection{Correction factor}
As previously discussed in Section 4, to estimate the LBG luminosity
function at $z=5$ and $z=6$ we used the $V/V_{max}$ maximum likelihood
estimator of Schmidt (1968), adapted to incorporate the photometric
redshift probability density function of each LBG
candidate. Consequently, within a given redshift interval
($z_{min}<z<z_{max}$), our estimate of the luminosity function within a given
absolute magnitude bin was:
\begin{equation}
\phi(M) = \sum_{i=1}^{i=N} \int^{z_{2}(m_i)}_{z_{1}(m_i)} 
\frac{C_{i}(m_i,z^{\prime})P_{i}(z^{\prime})\delta z^{\prime}}{V_{i}^{max}(m_i,z^{\prime})}
\end{equation}
\noindent
where the summation runs over the full sample, and $z_{1}\rightarrow z_{2}$ is
the redshift range within which a candidate's absolute magnitude lies
within the magnitude bin in question. As described in detail in
Section 4, the quantity $C_i(m_i,z)$ is a correction factor
derived from realistic monte-carlo simultions of the evolving LBG
population, and accounts for sample incompleteness, object blending
and contamination from objects photometrically scattered into the 
sample from faint-ward of the $z^{\prime}-$band magnitude limit.

To illustrate the influence of the correction
factor we show in Fig A1 our estimates of the $z=5$ and $z=6$ LBG luminosity
functions with and without the inclusion of $C_i(m_i,z)$. In
Fig A1 the open and filled black symbols are the estimates of the
$z=5$ and $z=6$ luminosity functions including the 
correction factor (identical to the estimates displayed in Fig 1),
while the grey symbols are the equivalent estimates setting
$C_i(m,z)=1$. It can be seen from Fig A1 that the correction factor 
does not have a significant impact on the shape of
the luminosity function estimates. At the bright end, where our sample
is most complete, the off-set between the black and grey data-points
is simply due to the $\simeq15$\% correction for objects lost due to
blending in the crowded Subaru images. In the faintest
absolute magnitude bins the inclusion of the correction factor boosts the
number densities by a roughly constant factor of $\simeq 30$\% (0.1 dex). 
This effect is due to a balance between the increasing positive
correction that has to be made for incompleteness as the data approach 
the $z^{\prime}=26$ magnitude limit, and the increasingly negative
correction that has to be made for objects photometrically scattered
into the sample from faint-ward of the magnitude limit. 

\subsection{Comparing different photometric redshift techniques}
When deriving our estimates of the LBG luminosity functions at $z=5$
and $z=6$ in Section 4 we used the full photometric redshift
probability density functions provided by our SED template fitting 
code in order to make full use of the available information. In this
section we investigate the influence of this technique on the derived 
luminosity function estimates. Of particular interest is to
investigate how the derived luminosity function estimates change if we 
simply treat the best-fitting photometric redshift for each LBG as
unique, and ignore the possibility of secondary redshift solutions.

\begin{figure}
\centerline{\psfig{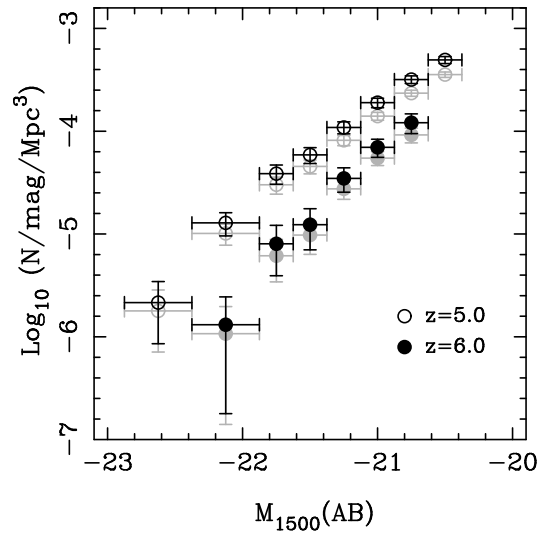}}
\caption{Estimates of the bright end of the UV-selected luminosity
function at $z=5$ and $z=6$. The open and filled black data-points are derived using Eqn A1,
including the correction factor $C_i(m_i,z)$, and are identical to
those presented in Fig 1. The grey data-points are the
corresponding estimates derived by setting $C_i(m_i,z)$=1 (see text
for discussion).}
\end{figure}

\begin{figure}
\centerline{\psfig{file=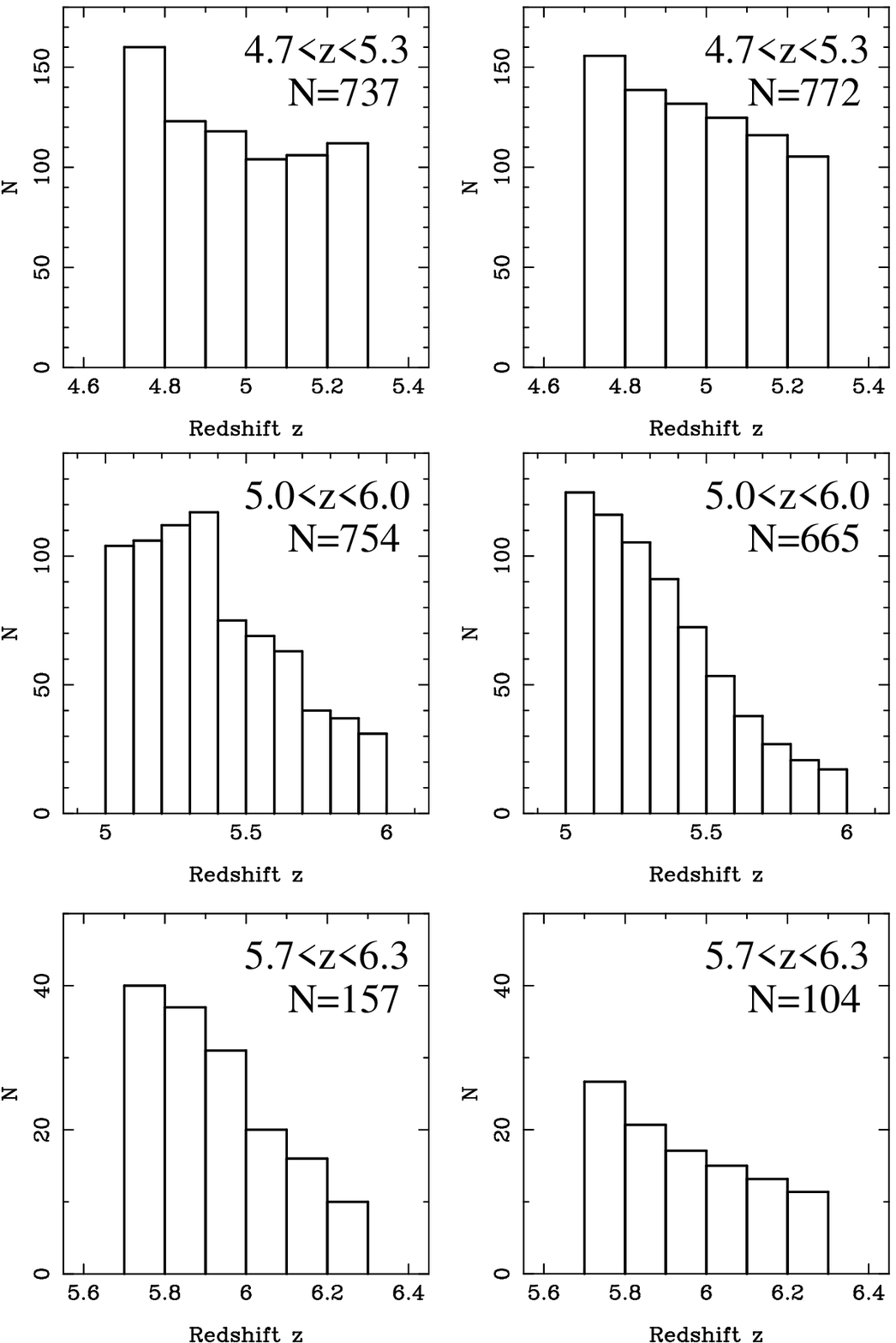,width=7.0cm,angle=0}}
\caption{A comparison of the LBG redshift distributions as derived using
the best-fitting photometric redshifts (left-hand column) and
integrating over the full redshift probability density function (right-hand column). 
The appropriate redshift range and the number of objects lying within
that range are listed in the top right-hand corner of each panel.}
\end{figure}

\subsubsection{Redshift distributions}
The first straightforward comparison of the two techniques is to
compare the estimated redshift distributions. In Fig A2 we show the
LBG redshift distributions in the three redshift intervals utilised in
this paper: $4.7<z<5.3$, $5.0<z<6.0$ and $5.7<z<6.3$. The redshift
distributions shown in the left-hand column are derived by treating
the best-fitting photometric redshift for each LBG as unique and
robust. The redshift distributions shown in the right-hand column are
derived by integrating over the full redshift probability density
functions as done in Section 4. The number of objects included in each
redshift distribution is listed in the top right-hand corner of each panel. 

Although the redshift distributions derived via both techniques are 
broadly similar, it can be seen from Fig A2 that they do differ in
detail. Perhaps the most noticeable difference is that adopting the 
best-fitting photometric redshifts alone predicts $\simeq50$\% more objects in
the highest redshift interval ($5.7<z<6.3$). Armed with this
information we now proceed to investigate how the differences in
redshift distributions affect the estimated luminosity functions.

\subsubsection{Luminosity function estimates}
In Fig A3 we show estimates for the bright end of the $z=5$ and
$z=6$ LBG luminosity functions. The left-hand panels show the estimates
derived in Section 4 by adopting the photometric redshift
probability density function for each LBG. The right-hand panels show
the luminosity function estimates derived from simply adopting the
best-fitting photometric redshift for each source. In each panel 
we also plot the maximum likelihood Schechter
function fits to the LBG luminosity function at either $z=5$ or $z=6$
as derived in Section 5.2.

It can be seen from Fig A3 that the estimate of the $z=5$ luminosity
function is virtually identical using both techniques. This is as
expected given the similarity of the $4.7<z<5.3$ redshift
distributions shown in the top panel of Fig A2, and demonstrates that
the photometric redshifts at $z\simeq5$ are sufficiently robust that
both techniques lead to indistinguishable conclusions. In contrast, although the two
luminosity function estimates at $z=6$ are clearly consistent, there
are differences in detail. Specifically, it can be seen from Fig A3
that the luminosity function estimate based on the best-fitting
photometric redshifts predicts more objects in the three faintest
absolute magnitude bins. This effect is entirely consistent with
the differences between the $5.7<z<6.3$ redshift distributions shown in
Fig A2, which demonstrate that using the best-fitting photometric
redshifts alone predicts $\simeq50$\% more objects at $z\simeq6$.

In conclusion, it is clear from the information displayed in Figs A2
\& A3 that the only substantive difference between the two techniques
is for the faintest LBGs at $z\simeq6$. This is perhaps to be
expected, given that these are the very objects for which the
best-fitting photometric redshifts are least robust, and was the
primary motivation for exploiting the extra information contained
within the full redshift probability density function for each
object. However, the good agreement between the luminosity function
estimate shown in the bottom right panel of Fig A3 and the
best-fitting Schechter function derived in Section 5.2 confirms that
adopting either technique would lead to the same conclusions regarding
the form, and evolution, of the LBG luminosity function between $z=5$
and $z=6$.

\begin{figure}
\centerline{\psfig{file=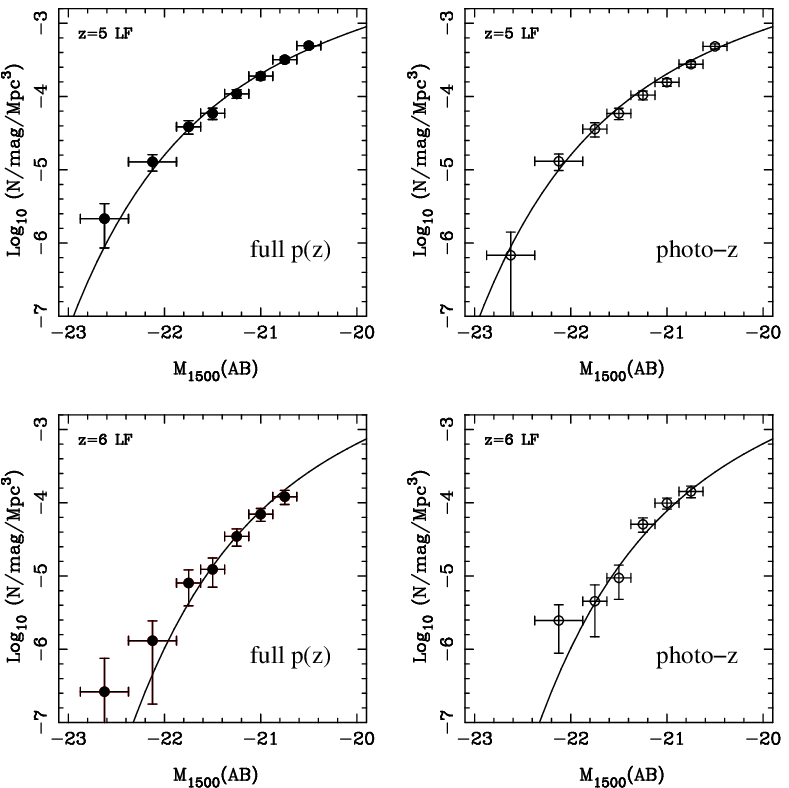,width=8.5cm,angle=0}}
\caption{A comparison of the LBG luminosity function estimates at
$z=5$ and $z=6$ resulting from using the full redshift probability
density function for each object (left-hand plots) and the
best-fitting photometric redshifts alone (right-hand plots). The top
two plots show the different luminosity function estimates at $z=5$,
while the bottom two plots show the different luminosity function
estimates at $z=6$. The maximum likelihood fits to the $z=5$ and $z=6$ luminosity functions
derived in Section 5.2 are plotted in the top and bottom panels respectively.}
\end{figure}

\end{appendix}
\end{document}